\documentclass[a4paper,review]{elsarticle}

\journal{}
\usepackage{psfrag}
\usepackage{subfigure}
\usepackage{amsthm}
\usepackage{amsmath}
\usepackage{amsfonts}
\usepackage{amssymb}
\usepackage{mathrsfs}
\usepackage{pbox}
\usepackage{enumitem}
\usepackage{url}
\usepackage[english]{babel}
\usepackage{color}
\usepackage{lineno,hyperref}
\usepackage{balance}
\newcommand{\virg}[1]{``#1''}
\DeclareMathOperator*{\argmin}{arg\,min}

\newtheorem{problem}{Problem}

\usepackage{xparse}%
\usepackage{blindtext}
\usepackage{morewrites}
\usepackage{wrapfig}
\usepackage{xkeyval}%

\newwrite\authorbibfile%

\AtBeginDocument{%
  \immediate\openout\authorbibfile=\jobname.aub%
}%

\AtEndDocument{%
\immediate\closeout\authorbibfile
\InputIfFileExists{\jobname.aub}{}{}
}%

\makeatletter

\define@key{authorbib}{scale}[1]{%
\def\AuthorbibKVMacroScale{#1}%
}

\define@key{authorbib}{wraplines}[10]{%
\def\AuthorbibKVMacroWraplines{#1}%
}

\define@key{authorbib}{imagewidth}[4cm]{%
\def\AuthorbibKVMacroImagewidth{#1}%
}

\define@key{authorbib}{overhang}[10pt]{%
\def\AuthorbibKVMacroOverhang{#1}%
}

\define@key{authorbib}{imagepos}[l]{%
\def\AuthorbibKVMacroImagepos{#1}%
}

\makeatother

\presetkeys{authorbib}{imagepos=l,imagewidth=4cm,wraplines=15,overhang=20pt}{}

\newlength{\AuthorbibTopSkip}
\newlength{\AuthorbibBottomSkip}
\NewDocumentCommand{\authorbibliography}{+o+m+m+m}{%
  \IfNoValueTF{#1}{%
  }{%
    \setkeys{authorbib}{#1}%
    \immediate\write\authorbibfile{%
      \string\begin{wrapfigure}[\AuthorbibKVMacroWraplines]{\AuthorbibKVMacroImagepos}[\AuthorbibKVMacroOverhang]{\AuthorbibKVMacroImagewidth}^^J
        \string\includegraphics[scale=\AuthorbibKVMacroScale]{#2}^^J
        \string\end{wrapfigure}^^J
    }%
  }%
  \IfNoValueTF{#3}{%
    \typeout{Warning: No author name}%
  }{%
    \immediate\write\authorbibfile{%
      \unexpanded{\vspace{\AuthorbibTopSkip}}^^J
      \string\noindent\relax
      \unexpanded{\textbf{#3}\par}^^J
      \string\noindent\relax
      \unexpanded{#4}^^J%
      \unexpanded{\vspace{\AuthorbibBottomSkip}}^^J
      }%
  }%
}%
\begin{document}

\begin{frontmatter}

\title{An Adaptive Groundtrack Maintenance Scheme\\ for Spacecraft with Electric Propulsion}
\author{Mirko Leomanni\fnref{myfootnote}}
\ead{leomanni@diism.unisi.it}
\fntext[myfootnote]{Corresponding author}
\author{Andrea Garulli}\ead{garulli@diism.unisi.it}
\author{Antonio Giannitrapani}\ead{giannitrapani@diism.unisi.it}
\address{Dipartimento di Ingegneria dell'Informazione e Scienze Matematiche\\ Universit\`a di Siena, Siena, Italy}
\author{Fabrizio Scortecci}\ead{fscortecci@aerospazio.com}
\address{Aerospazio Tenologie\\ Rapolano Terme, Siena, Italy}

\begin{abstract}
In this paper, the repeat-groundtrack orbit maintenance problem is addressed for spacecraft driven by electric propulsion. An adaptive solution is proposed, which combines an hysteresis
controller and a recursive least squares filter. The controller provides a pulse-width modulated command to the thruster, in compliance with the peculiarities of the electric propulsion technology. The filter takes care of estimating a set of environmental disturbance parameters, from inertial position and velocity measurements. The resulting control scheme is able to compensate for the groundtrack drift due to atmospheric drag, in a fully autonomous manner. A numerical study of a low Earth orbit mission confirms the effectiveness of the proposed method.
\end{abstract}


\end{frontmatter}

\section{Introduction}
Recent years have seen a growing trend in the development of Low Earth Orbit (LEO) space missions for commercial, strategic, and scientific purposes. Examples include Earth Observation programs such as DMC, RapidEye  and Pl\'eiades-Neo \cite{dmcmission,tyc2005rapideye}, as well as broadband constellations like OneWeb, Starlink and Telesat \cite{mclain2017future}.  As opposed to geostationary spacecraft, a LEO satellite cannot stay always pointed towards a fixed spot on Earth. Nevertheless, it can revisit the same location periodically, by using a repeat-groundtrack orbit configuration. Besides its utility in traditional remote sensing applications, such a configuration can be exploited to deploy satellite constellations supporting global, regional and reconnaissance services \cite{mortari2004flower,ruggieri2006flower}. Thus, a repeat-groundtrack orbit design is ubiquitous in the above-mentioned type of missions.

Due to environmental perturbations and, in particular, to atmospheric drag, the groundtrack of a LEO satellite tends to drift away from the nominal repeat condition. This undesired effect must be compensated for by means of a suitable groundtrack maintenance program. Depending on the required control accuracy, the program may involve a tight maneuvering schedule, which makes ground-based control inefficient and risky. In order to overcome this issue, a variety of autonomous on-board control strategies have been proposed, see e.g., \cite{wertz1998autonomous,garulli2011autonomous,de2014virtual,weiss2018station}.

Groundtrack maintenance operations require an adequately sized propulsion unit. The unit must produce the delta-v associated to the maintenance program, in addition to the one reserved for orbit acquisition and de-orbiting maneuvers. The latter, in particular, are mandatory within present regulations \cite{inter2002iadc}. For very low altitude orbits, drag compensation can be the dominant factor in the mission delta-v budget, while the contribution due to de-orbiting becomes more pronounced at higher altitudes. In any case, the total delta-v may grow to a level justifying the adoption of a high specific impulse, low-thrust technology such as Electric Propulsion (EP) \cite{guelman1999electric,fearn2005economical,Wright201548,leomanni2017propulsion}. Indeed, a number of EP-based LEO missions have been recently launched \cite{lev2017technological}.

The maintenance of repeat-groundtrack orbits is a well-established topic in astrodynamics. The classical open-loop solution to the maintenance problem is described, for instance, in \cite{mag11}. Several modifications to this method have been proposed in the literature. In  \cite{aorpimai2007repeat}, a feedback implementation is presented. The effect of a moderate orbital eccentricity is analyzed in \cite{sengupta2010satellite}. The application to successive-coverage orbits is discussed in \cite{fu2012design}. A high-precision design is presented in \cite{he2017high}.
In all these works, each engine burn is modeled as an impulsive velocity change. Such an approximation is not well suited for EP engines, whose thrust profile is usually modeled as a rectangular pulse. In light of this consideration, some recent studies \cite{zhang2016coplanar,di2018continuous} have started adopting a piecewise-constant parametrization of the control command.

In this paper, an autonomous groundtrack maintenance strategy is developed for EP-based LEO missions. In compliance with the peculiarities of the EP technology, it is assumed that the thruster is either operated at a constant set-point or switched off, thus requiring a pulse-width modulated control input signal. It is shown that, for small deviations about a circular orbit, the groundtrack error dynamics is that of a perturbed double integrator. The perturbation term is modeled as a periodic signal with non-zero mean, according to what is observed in the literature \cite{mag11}. Within this setting, an adaptive feedback scheme is derived. The idea underpinning the design is to combine an hysteresis controller with a recursive least squares (RLS) filter. The RLS filter is in charge of estimating a set of parameters of the error system. The controller provides a suitable on/off engine switching signal based on the filter estimates. The control law builds upon recent results on minimum switching control in \cite{garulli2015minimum,leomanni2017minimum}.

The proposed control scheme is evaluated on a simulation case study featuring a 460 km altitude orbit. Simulation results show that the desired repeat-groundtrack pattern is acquired successfully and that the tracking error is maintained within prescribed limits, relying only on GPS measurements.

The rest of the paper is organized as follows. In Section \ref{sec2}, the repeat-groundtrack control problem is reviewed. Section \ref{sec3} presents the dynamic model which is employed for control design. The adaptive groundtrack control scheme is described in Section \ref{sec4}. The simulation case study is discussed in Section \ref{sec5}, and conclusions are drawn in Section \ref{sec6}.

\section{Problem Formulation}\label{sec2}
Repeat-groundtrack orbits depend on the commensurability between the satellite nodal period (the time interval it takes a satellite to make two consecutive ascending node crossings), and the nodal period of Greenwich (the period of the Earth's rotation with respect to the ascending node). The nodal period is defined as
\begin{equation}\label{Tgamma}
T_\gamma=\frac{2\pi}{\dot{\gamma}},
\end{equation}
where $\gamma=M+\omega$ is the satellite mean latitude, i.e., the sum of the mean anomaly $M$ and of the argument of periapsis $\omega$.
The nodal period of Greenwich is given by
\begin{equation}\label{TG}
T_G=\frac{2\pi}{{\omega}_{\oplus}-\dot{\Omega}},
\end{equation}
where ${\omega}_{\oplus}$ is the Earth's rotation rate, and $\Omega$ is the satellite right ascension of the ascending node.
Thus, the repeat-groundtrack  condition can be formalized as
\begin{equation}\label{rcon}
T_\gamma=r \;{T_G},
\end{equation}
where $r>0$ is a rational number, representing the ratio between the number of days and the number of satellite revolutions within the repeat cycle. For LEO orbits, $r$ is typically in the order of $1/15$.

The groundtrack spacing $\lambda_S$ between two consecutive ascending node crossings is given by
\begin{equation}\label{lambdaS}
\lambda_S=2\pi\, \frac{T_\gamma}{T_G}.
\end{equation}
Ideally, $\lambda_S=2\pi r$.  However, in the presence of perturbations such as atmospheric drag and third-body gravity, $\lambda_S$ drifts away from the nominal value. In standard repeat-groundtrack control schemes, a sequence of impulsive orbit adjustment maneuvers is commanded to counteract this change \cite{mag11}. In this paper, a different approach is proposed.

Besides enforcing \eqref{rcon}, the control scheme must guarantee that the satellite repeatedly crosses the equator at a desired longitude $\lambda^*\in[0,2\pi)$, east of Greenwich. Let $\lambda_G(t)=\lambda_G(0)+{\omega}_{\oplus}t$ be the instantaneous longitude of Greenwich. Then, the groundtrack error $x(t)$ is defined as
\begin{equation}\label{trkerr}
x(t):=r\,\gamma(t)+\,\Omega(t)-\lambda_G(t)-\lambda^*,
\end{equation}
where all angular quantities are unwrapped. Notice that when $x(t)$ is constant, i.e.\ $\dot{x}(t)=0$, one has
\begin{equation}\label{rcon1}
r\,\dot{\gamma}+\dot{\Omega}-{\omega}_{\oplus}=0,
\end{equation}
which in turn, by using \eqref{Tgamma} and \eqref{TG}, implies that \eqref{rcon} is satisfied. Moreover, if  $x(t)=0$, then the satellite crosses the equator at the desired east longitude $\lambda^*$. In fact, by definition
$
\gamma=2 \pi N,
$
with $N\in\mathbb{N}$, whenever the satellite crosses the ascending node. Being $r$ in \eqref{rcon} a rational number, there exists an ascending node crossing time $t_c$, at which
\begin{equation}\label{modr}
\left[r\,{\gamma(t_c)}\right]\;\text{mod}\, 2\pi=0.
\end{equation}
By using \eqref{modr} and $x(t)=0$, it follows from \eqref{trkerr} that
\begin{equation}
[\Omega(t_c)-\lambda_G(t_c)]\;\text{mod}\, 2\pi=\lambda^*,
\end{equation}
where the expression on the left hand side is indeed the satellite east longitude at the equator crossing time $t_c$.

The requirement $x(t)=0$, $\forall t$, cannot be achieved with a on/off thrusting strategy. Therefore, the following relaxed control problem is addressed.
\begin{problem}\label{pb1}
Find an on/off thrusting scheme, ensuring that
\begin{equation}\label{obj}
| x(t) | \leq x_\text{lim} \quad \text{for all}\;t\geq\tilde{t}>0,
\end{equation}
where  $x_\text{lim}>0$ is a predefined longitude error tolerance, and $\tilde{t}$ is a finite settling time.
\end{problem}

Following a common practice, we consider tangential thrust only. In fact, radial and out-of-plane maneuvers are often deemed too expensive in terms of fuel consumption \cite{mag11}. In this regard, it should be noticed that Problem \ref{pb1} does not account for the grountrack deviation resulting from inclination errors (which is zero at the equator and increases with the latitude). Such contribution can be compensated for separately, if required, via out-of-plane maneuvers.
Moreover, attention is restricted to the case of a near-circular orbit, since most LEO spacecraft are flown in this type of orbit.

\section{Groundtrack Error Dynamics}\label{sec3}
The dynamics of the groundtrack error are obtained by taking the second derivative of \eqref{trkerr} with respect to time, resulting in
\begin{equation}\label{sysdyn}
\ddot{x}(t)=r\,\ddot{\gamma}(t)+ \ddot{\Omega}(t).
\end{equation}
Let the scalar $u_T(t)$ be a tangential control acceleration and the vector $d(t)=[d_R\; d_T\; d_N]^T$ describe the radial, tangential and normal components of a perturbation due to environmental sources. The effect of $u_T$ and $d$ on system \eqref{sysdyn} can be modeled through the following variational equations, adapted for nearly circular orbits (see, e.g., \cite{garulli2011autonomous,d2006proximity})
\begin{equation}\label{vareq}
\begin{array}{l l l}
\dfrac{\text{d}}{\text{dt}}\,\gamma&=&n-\sqrt{\dfrac{a}{\mu}}\Big[2 d_R +\sin(\gamma)\cot(i)\,d_N\Big]
\\[3mm]
\dfrac{\text{d}}{\text{dt}}\,n&=&-\dfrac{3}{a}(d_T+u_T)
\\[3mm]
\dfrac{\text{d}}{\text{dt}}\,\Omega&=&\sqrt{\dfrac{a}{\mu}}\,\dfrac{\sin(\gamma)}{\sin(i)}\, d_N
\\[3mm]
\dfrac{\text{d}}{\text{dt}}\, i&=&\sqrt{\dfrac{a}{\mu}}\,{\cos(\gamma)}\, d_N,
\end{array}
\end{equation}
where the dependence on time is left implicit, $n=\sqrt{\mu/a^3}$ denotes the mean motion, $a$ and $i$ indicate the semi-major axis and inclination, respectively, and $\mu$ is the gravitational parameter.  By using \eqref{vareq}, it can be verified that, for small deviations about an orbit with semi-major axis $a^*$, \eqref{sysdyn} takes on the form
\begin{equation}\label{sysdyn1}
\ddot{x}(t)=p_d(t)-\frac{3\,r}{a^*}\,u_T(t),
\end{equation}
where the time varying quantity $p_d(t)$ describes the cumulative effect of environmental perturbations.

Finding an analytical expression for $p_d(t)$ is in general a formidable task. Moreover, such an expression will unavoidably suffer form inaccuracies in the perturbation model adopted for the term $d(t)$. For the problem at hand, it is known that $p_d(t)$ is approximately constant, on average, over few orbital periods \cite{mag11}. Hence, within such a time scale, $p_d(t)$ can be modeled as
\begin{equation}\label{disturbance}
p_d(t)= p +  p_\Delta(t),
\end{equation}
where ${p}$ denotes the average contribution of perturbations, and the term $p_\Delta(t)$ accounts for zero-mean periodic effects.

It is worth noticing that the tangential acceleration due to atmospheric drag usually represents the major contribution to the disturbance component $p$ in \eqref{disturbance}. More specifically, one has that $p\simeq-{3\,r\, \bar{d}_T}/{a^*}>0$, where $\bar{d}_T<0$ denotes the average drag acceleration.
Conversely, secular perturbations on $\gamma$ and $\Omega$ have a limited impact on \eqref{disturbance}. For instance, a secular drift of $\Omega$ due to the Sun-synchronicity condition $\dot{\Omega}=2\pi$ rad/year would result in $\ddot{\Omega}=0$ in \eqref{sysdyn}, which in turn does not provide any contribution to \eqref{disturbance}.

\section{Repeat-Groundtrack Control Scheme}\label{sec4}
In this Section, a on/off control law is presented for Problem \ref{pb1}, under the simplifying assumptions that full state information is available and that $p_\Delta(t)=0$ in \eqref{disturbance}. Then, a RLS filter is derived, which estimates both the system state and the average disturbance $p$. Finally, the implementation of an adaptive control scheme based on this modules is discussed, together with a tuning strategy which minimizes the variation of the orbital eccentricity due to thrusting.
\subsection{Controller Design}\label{sec4a}
In order to compensate for a constant positive disturbance $p_d(t)=p$ in \eqref{sysdyn1}, consider a on/off control input of the form
\begin{equation}\label{swinp}
u_T=u_\text{max}\, v(t),
\end{equation}
where $u_{\text{max}}>0$ is the maximum acceleration which can be delivered by the propulsion system and $v(t)\in\{0,1\}$ is the engine activation signal. By enforcing \eqref{swinp} and $p_\Delta(t)=0$ in \eqref{sysdyn1}-\eqref{disturbance}, the following averaged model is obtained
\begin{equation}\label{sysdyn2}
\ddot{y}(t)=p-k\, v(t),
\end{equation}
where
\begin{equation}\label{kpar}
k=\frac{3\,r\,u_{\text{max}}}{a^*},
\end{equation}
and $y(t)$ denotes the average groundtrack error corresponding to $p_\Delta(t)=0$.

Let $y(t)$ be the solution to system \eqref{sysdyn2} starting from the initial conditions $y(0)=x(0)$ and $\dot{y}(0)=\dot{x}(0)$. Clearly, the relationship between $y(t)$ and the solution $x(t)$ of system \eqref{sysdyn1} is
\begin{equation}\label{difffyn}
y(t)=x(t)-\alpha(t),
\end{equation}
where
\begin{equation}\label{beta}
\alpha(t)=\textstyle \iint_{\,0}^{\,t} p_\Delta(\tau)\; \text{d}\tau
\end{equation}
is a zero-mean periodic signal. By using \eqref{sysdyn2}-\eqref{difffyn}, Problem \ref{pb1} can be recast as that of finding a switching signal $v(t)$, guaranteeing that
\begin{equation}\label{obj2}
| y(t) | \leq y_\text{lim} \quad \text{for all}\;t\geq\bar{t}>0,
\end{equation}
where
\begin{equation}\label{limcondition}
0<y_\text{lim}\leq x_\text{lim}-\max_t |\alpha(t)|.
\end{equation}
Notice from \eqref{sysdyn2} that the condition $k>p$ must be met, in order for the problem to be solvable. Moreover, by \eqref{limcondition}, one must have $x_\text{lim}>\max_t |\alpha(t)|$,
which is typically the case in real-world applications  {(since high-frequency disturbances $p_\Delta(t)$ are attenuated by the double integrator system \eqref{sysdyn1}-\eqref{disturbance})}.

\begin{figure}[!t]
\centering
\psfrag{a}{$y$}
\psfrag{b}{$\dot{y}$}
\psfrag{d}{$\textcolor{blue}{s(y,\dot{y};p)=\!-y_\text{lim}}$}
\psfrag{c}{$\textcolor{blue}{s(y,\dot{y};p)=y_\text{lim}}$}
\psfrag{e}{$y_\text{lim}$}
\psfrag{f}{$\hspace{-2mm}-y_\text{lim}$}
\includegraphics[width=0.85\linewidth]{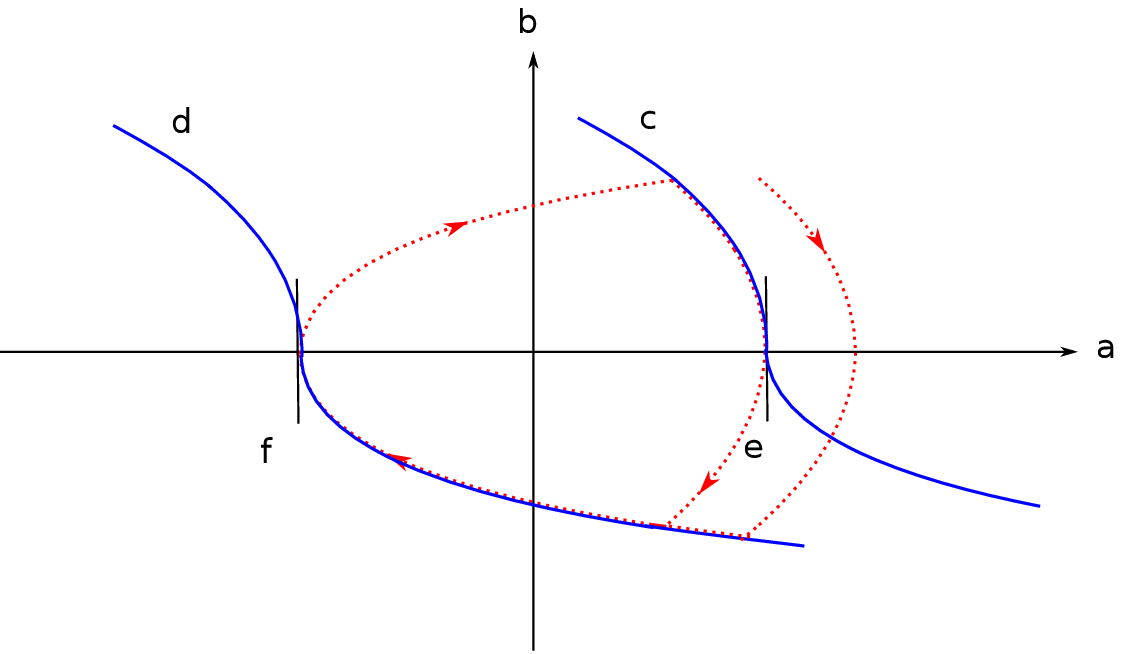}
\caption{Switching curves (solid) and example of a trajectory (dotted) resulting from the application of the control scheme \eqref{sfopt3}.}
\label{ovetto}
\end{figure}
A minimum fuel and minimum switching solution to the above control problem is obtained by exploiting the results in \cite{garulli2015minimum}. Consider the switching function
\begin{equation}\label{scurves}
s(y,\dot{y};p)=\left\{
\begin{array}{l l}
y-\dfrac{1}{2(p-k)}\,\dot{y}^2 & \;\text{if}\quad  \dot{y} \geq 0 \vspace{1mm}\\
y-\dfrac{1}{2 p }\,\dot{y}^2   & \;\text{if}\quad  \dot{y} < 0 .
\end{array}\right.
\end{equation}
Let the on/off input $v(t)$ be specified according to the hysteresis function
\begin{equation}\label{sfopt3}
v(t)=\left\{
\begin{array}{c  l}
 1 &\text{if}\quad  s(y(t),\dot{y}(t);p) \geq \, y_\text{lim}\\
 0 &\text{if}\quad          s(y(t),\dot{y}(t);p) \leq -y_\text{lim}\\
 v_h & \text{otherwise},
\end{array}\right.
\end{equation}
where $v_h=1$ if $s(y,\dot{y};p)\geq y_\text{lim}$ occurred more recently than $s(y,\dot{y};p) \leq \!-y_\text{lim}$, $v_h=0$ otherwise.
By applying the control law \eqref{sfopt3} to system \eqref{sysdyn2}, a limit cycle trajectory with amplitude $2 y_\text{lim}$ is reached in finite time from any initial condition. An example is shown in the phase plane portrait reported in Fig. \ref{ovetto}. The limit cycle period is given by $T_L=4\sqrt{\frac{k\,y_\text{lim}}{p\,k-p^2}}$. It is divided into a firing period of length $T_F=D T_L$ and a coasting period of length $T_C=(1-D) T_L$, where $D=p/k$ denotes the actuator duty cycle. Thus, the firing time turns out to be
\begin{equation}\label{firetime}
T_F=\frac{4p}{k}\sqrt{\frac{k \, y_\text{lim}}{p\,k-p^2}}=4\sqrt{\frac{p\, y_\text{lim}}{k^2-k\, p}}.
\end{equation}

\subsection{RLS Filter Design}
The control law \eqref{sfopt3} requires the real-time knowledge of $y(t)$, $\dot{y}(t)$ and $p$. Given the parameter $p$ and the initial conditions $y(0)$, $\dot{y}(0)$, one can compute $y(t)$ and $\dot{y}(t)$ just by integrating system \eqref{sysdyn2}. However, using this solution would result in an open-loop control strategy. Moreover, the exact value of $y(0)$, $\dot{y}(0)$ and $p$ is, in general, unknown.

An alternative approach consists in estimating $y(t)$, $\dot{y}(t)$ and $p$ by using values of $x(t)$ obtained from inertial measurements. More specifically, the orbital elements $\gamma(t)$ and $\Omega(t)$ in \eqref{trkerr} can be computed from absolute position and velocity measurements, by using standard analytical methods \cite{mag11}, while the quantities $\lambda_G(t)$, $r$ and $\lambda^*$ are known.

Let the obtained values of $x(t)$ be denoted as $\tilde{x}(t_j)$, where $\{t_j\}_{j\in\mathbb{N}}$ is the sequence of time samples at which measurements are taken. Now, observe that the solution to \eqref{sysdyn2}, with either $v=0$ or $v=1$, can be parameterized as
\begin{equation}\label{parsol}
y(t)=\varphi^T(t){\theta},
\end{equation}
where $\varphi(t)=[t^2\;t\;\,1]^T$ and ${\theta}=[{\theta}_1\;{\theta}_2\;{\theta}_3]^T$ is a vector of unknown parameters to be determined. Hence, a least squares estimation problem can be cast as follows
\begin{equation}\label{difffyn2}
\hat{\theta}=\underset{{\theta}}{\argmin} \sum_{j} \, [\,\tilde{x}(t_j)-\varphi^T(t_j)\,{\theta}\,]^2.
\end{equation}
 A recursive solution to \eqref{difffyn2} is provided by the RLS algorithm \cite{ljung1998system}
\begin{eqnarray}
\hat{\theta}(t_j)&=&\hat{\theta}(t_{j-1})+\frac{P(t_{j-1})\,\varphi(t_j)}{\eta+\varphi(t_j)^T P(t_{j-1})\,\varphi^(t_j)}\left[\tilde{x}(t_j)-\varphi^T(t_j)\hat{\theta}(t_{j-1}) \right]
\label{thupd}\\[2mm]
P(t_j)&=&\frac{1}{\eta}\left[P(t_{j-1})-\frac{P(t_{j-1})\,\varphi(t_j)\varphi^T(t_j)\,P(t_{j-1})}{\eta+\varphi^T(t_j)\,P(t_{j-1})\,\varphi(t_j)}\right], \label{Pupd}
\end{eqnarray}
initialized at $P(t_{0})$ and $\hat{\theta}(t_0)$, where $\eta$ is the forgetting factor.

The algorithm \eqref{thupd}-\eqref{Pupd} is re-initialized each time that the right hand side of \eqref{sysdyn2} changes sign, i.e., whenever the input switches, in order to prevent divergence in the estimates. Let $\bar{t}$ be an input switching time, such that $t_{s-1}< \bar{t}\leq t_{s}$. At time $\bar{t}$, the parameter vector is reset according to
\begin{equation}\label{threset}
\begin{array}{l l l}
\hat{\theta}_3(\bar{t})&=&\varphi^T(\bar{t})\,\hat{\theta}(t_{s-1})\\
\hat{\theta}_2(\bar{t})&=&2\, \hat{\theta}_1(t_{s-1})\, \bar{t} +\hat{\theta}_2(t_{s-1})\\
\hat{\theta}_1(\bar{t})&=&\hat{\theta}_1(t_{s-1}) +k[v(t_{s-1}) -\,v(\bar{t})]/2.
\end{array}
\end{equation}
Then, \eqref{thupd}-\eqref{Pupd} is applied by replacing $\varphi(t_j)$ with $\varphi(t_j-\bar{t})$, for $t_j\geq t_s$, and by setting the initial condition to $\hat{\theta}(\bar{t})$ given by \eqref{threset}. This ensures the continuity of $\varphi^T\hat{\theta}$ and of its first time derivative.

%\begin{figure}[!t]
%\centering
%\includegraphics[width=0.65\linewidth]{filex.eps}
%\caption{Profiles of the signals $x(t)$ and $\hat{y}(t)$ for an example simulation with $v=0$.}
%\label{filex}
%\end{figure}
The continuous-time estimates of  $y(t)$, $\dot{y}(t)$ and $p$ returned by the filter are obtained by interpolation, as follows
\begin{equation}\label{pest}
\begin{array}{l l l}
\hat{y}(t)&=& \varphi^T\!(t-\bar{t})\,\hat{\theta}(t_j)  \\
\hat{\dot{y}}(t)&=&2(t-\bar{t})\,\hat{\theta}_1(t_j)+\hat{\theta}_2(t_j)\\
\hat{p}(t)&=&2\, \hat{\theta}_1(t_j)+k \,v(t),
\end{array}
\end{equation}
where $t_j$ and $\bar{t}$ denotes the most recent measurement time and the most recent input switching time, respectively.

%{Notice that the forgetting factor $\eta$ in \eqref{thupd}-\eqref{Pupd} provides the %ability to adapt the filter estimates \eqref{pest} to low-frequency variations in ${p}_d(t)$, %which are not modeled by \eqref{disturbance}.}

%Figure \ref{filex} shows the signals $x(t)$ and $\hat{y}(t)$ for an example simulation with %$v=0$.

\subsection{Adaptive Control Scheme Implementation}\label{ACS}
\begin{figure}[!t]
\centering
\psfrag{d}{$p_d$}
\psfrag{p}{$\hat{y},\,\hat{\dot{y}},\,\hat{p}$}
\psfrag{u}{$v$}
\psfrag{r}{$x$}
\psfrag{y}{$y_\text{lim}$}
\includegraphics[width=0.8\linewidth]{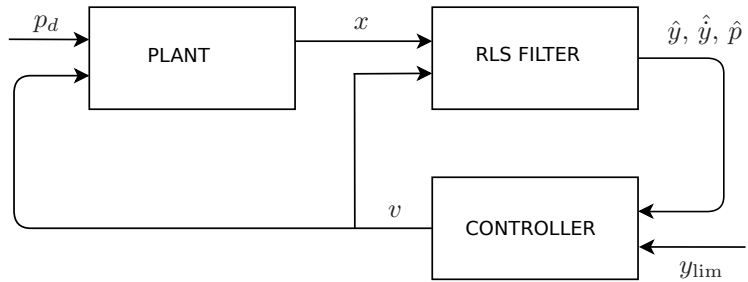}
\caption{Block diagram of the closed-loop system.}
\label{blockdiag}
\end{figure}

A real-time feedback control scheme is obtained by replacing $y(t)$, $\dot{y}(t)$ and $p$ in \eqref{sfopt3}, with the corresponding estimates $\hat{y}(t)$, $\hat{\dot{y}}(t)$ and $\hat{p}(t)$ provided by \eqref{pest}. The resulting adaptive control system is shown in Fig.~\ref{blockdiag}.
%{ It is based on the assumption that $\hat{p}(t)$ does not change significantly during the limit cycle period $T_L$, which is dictated by the proposed controller design.}

It should now be remarked that the relationship \eqref{limcondition} leaves some freedom in the choice of the control specification $y_\text{lim}$. Such degree of freedom can be exploited to optimize other relevant performance criteria. Specifically, some applications require the satellite altitude to vary as little as possible, which corresponds to maintaining a near-zero orbital eccentricity. In this regard, atmospheric drag is know to have a stabilizing effect, as it tends to circularize the orbit. However, repeated groundtrack adjustment maneuvers can lead to a secular eccentricity growth. For nearly circular orbits, the variation in the orbital eccentricity due to the control acceleration $u_T$ can be modeled as \cite{fu2012design}
\begin{equation}\label{decc}
\dot{e}= 2\sqrt{\dfrac{a}{\mu}}\cos(f)\, u_T,
\end{equation}
where $e$ indicates the orbital eccentricity and $f$ denotes the true anomaly. For small deviations about $a=a^*$, equation \eqref{decc} can be approximated as
\begin{equation}\label{decc2}
\dot{e} \approx 2\sqrt{\dfrac{a^*}{\mu}}\cos(n^* t+ f_0)\, u_T,
\end{equation}
where $n^*=\sqrt{\mu/(a^*)^3}$ is the reference mean motion.

The net change in $e$, $\Delta e=e(T)-e(0)$, obtained by applying a constant input $u_T=u_\text{max}$ over a time interval of length $T$, can be computed from \eqref{decc2} according to
\begin{equation}\label{decc3}
\Delta{e}=2\,u_\text{max}\sqrt{\dfrac{a^*}{\mu}}\!\int_0^T\!\cos(n^* t+ f_0)\,\text{d}t=\dfrac{2 (a^*)^2u_\text{max}}{\mu}\Big[\sin(n^* T+ f_0)-\sin(f_0) \Big].
\end{equation}
From \eqref{decc3} it follows that one possibility  to enforce a zero secular growth $\Delta{e}=0$ of the eccentricity is to set $T=T(m)=m\,\frac{2\pi}{n^*}$, i.e. to fire the engine for an integer multiple $m$ of the orbital period. Another possibility is to initiate the maneuver at $f_0=0$ or $f_0=\pi$ (assuming the periapsis exists) and set $T(m)=m\,\frac{\pi}{n^*}$, which amounts to firing for a multiple of half the orbital period (in particular, $T=\pi/{n^*}$ may be adopted to avoid firing the engine during eclipses). The corresponding $y_\text{lim}$ can be found by equating $T(m)$ to the firing time $T_F$ in \eqref{firetime}, thus resulting in
\begin{equation}\label{ylimst}
y_\text{lim}=\frac{k(k-p)}{16\,p} \,T^2(m).
\end{equation}
In order for the above strategy to be feasible, $y_\text{lim}$ must satisfy condition \eqref{limcondition}. For typical values of $x_\text{lim}$, $k$ and $p$, this requirement can be met by adopting a sufficiently small value of $m$ in \eqref{ylimst}. In this work, \eqref{ylimst} is evaluated on-line, by using the disturbance estimate $\hat{p}$ returned by the RLS filter.

Finally, notice that the control law \eqref{scurves}-\eqref{sfopt3} by itself does not guarantee the firing sequence to be initiated at a specific true anomaly location $f_0$. In order to do so, once the condition $s({y},\dot{y};p)\geq y_\text{lim}$ is met, one has to postpone the engine activation $v=1$ until $f$ reaches $f_0$. The groundtrack error will grow only slightly during this time interval, since the true longitude dynamics are usually much faster than the error ones.

%(where m can be set to 0 or 1 according to the perigee location)
\section{Simulation Case Study}\label{sec5}
An Earth observation mission performed by a 200 kg minisatellite equipped with EP has been simulated numerically, in order to validate the proposed approach. The spacecraft bus layout is modeled as a cube with side-length equal to 1 m. The truth model for the simulation consists of a numerical propagation routine based on Cowell's method, which accounts  for the most relevant environmental perturbations affecting LEO satellites. Table \ref{propmod} describes the main features of the simulation environment.
\begin{table}[h]
\caption{Main features of the simulation model}\label{propmod}
\centering\vspace{3mm}
\begin{tabular}{| l | l |}
%\hline
\hline
\multicolumn{1}{|c|}{Contribution} & \multicolumn{1}{c|}{Model} \\ \hline
Earth's Gravity &     EGM96 $9 \times 9$   \\
Atmospheric Drag &     NRLMSISE-00, $F_{10.7}=220$, $A_{p}=15$ \\
Third Body &    Luni-solar point mass gravity   \\
Solar Pressure  &    Cannonball model with eclipses
\\ \hline
\end{tabular}
\end{table}

The satellite is released in a near-circular sun-synchronous orbit with an altitude of about 460 km. The initial orbital elements, reported in Table \ref{orbelsim}, are chosen so as to achieve a repeat-groundtrack  period of 3 days ($r=3/46$ days/revs). Hence, the reference semi-major axis in \eqref{sysdyn1} is $a^*=a(0)$. It is required to keep groundtrack within a maximum deviation of 2 km from the nominal one at the equator, which corresponds to the error tolerance $x_\text{lim}=2/R_\oplus=3.136\cdot 10^{-4}$ rad, where $R_\oplus$ denotes the Earth's equatorial radius. This level of accuracy is compatible with advanced scientific missions such as ICEsat \cite{schutz2005overview}.
\begin{table}[t]
\caption{Initial conditions for the simulation}\label{orbelsim}
\centering\vspace{3mm}
\begin{tabular}{| l | l |}
%\hline
\hline
\multicolumn{1}{|c|}{Orbital element}  & \multicolumn{1}{c|}{Initial value} \\ \hline
Semi-major axis &     $a(0)=6838$ km    \\
Eccentricity &    $e(0)=0.001$    \\
Inclination &     $i(0)=97.28$ deg  \\
RAAN  &    $\Omega(0)=0$ deg    \\
Argument of periapsis  &   $\omega(0)=90$ deg  \\
True anomaly   &  $f(0)=270$ deg
\\ \hline
\end{tabular}
\end{table}
\begin{table}[b]
\caption{Characteristics of the control system}\label{gpshet}
\centering\vspace{3mm}
\begin{tabular}{| l | l | l | c |}
%\hline
\hline
\multicolumn{1}{|c|}{Device}  & \multicolumn{1}{c|}{Output}  & \multicolumn{1}{c|}{Std dev} & \multicolumn{1}{c|}{Update time}\\ \hline
GPS &
\begin{tabular}{l}
{Inertial position}\\
{Inertial velocity}
\end{tabular}
&
\begin{tabular}{l}
20 m\\
0.1 m/s
\end{tabular}
& $30$ sec \\ \hline
HET &
\begin{tabular}{l}
{ON: $10$ mN thrust}\\
{OFF: no thrust}
\end{tabular}
&
\begin{tabular}{c}
{$0.5$ mN }\\
{-}
\end{tabular}
& $30$ sec
\\ \hline
\end{tabular}
\end{table}

The groundtrack control system relies on a GPS receiver providing position and velocity measurements, and on a low-power Hall Effect Thruster (HET). Table \ref{gpshet} reports the characteristics of the model used for the GPS and the HET. It is assumed that the thruster is aligned with the direction tangential to the orbit. The nominal acceleration provided by the HET amounts to $u_\text{max}=F_\text{max}/m_{\text{sat}}=5\cdot 10^{-5}$  m/s$^2$, where $F_\text{max}=0.01$ N and $m_\text{sat}=200$ kg.

The measured values $\tilde{x}(t_j)$ of $x(t)$ are computed by first expressing GPS position and velocity measurements in terms of osculating orbital elements, and then transforming osculating elements into mean ones according to \cite{brouwer1959solution}.
The forgetting factor of the RLS algorithm \eqref{thupd}-\eqref{Pupd} is set to $\eta=0.9999$. This provides a good trade-off between the sensitivity of the filter to high-frequency oscillations in $x(t)$ and its responsiveness to long-period drifts in $p_d(t)$ (e.g., due to seasonal changes in the atmospheric density), which are not modeled by \eqref{disturbance}. The error tolerance $y_\text{\text{lim}}$ in \eqref{obj2} is set according to \eqref{ylimst}, with $T=T(m)=2\pi/n^*$ and $p=\hat{p}$, resulting in $y_\text{\text{lim}}\simeq 2\cdot 10^{-4}$ rad. This ensures that condition \eqref{limcondition} is met with a good safety margin; at the same time, it allows one to successfully counteract the eccentricity variation in \eqref{decc3}, according to the discussion in Section \ref{ACS}.

\begin{figure}[t]
\centering
\includegraphics[width=0.85\linewidth]{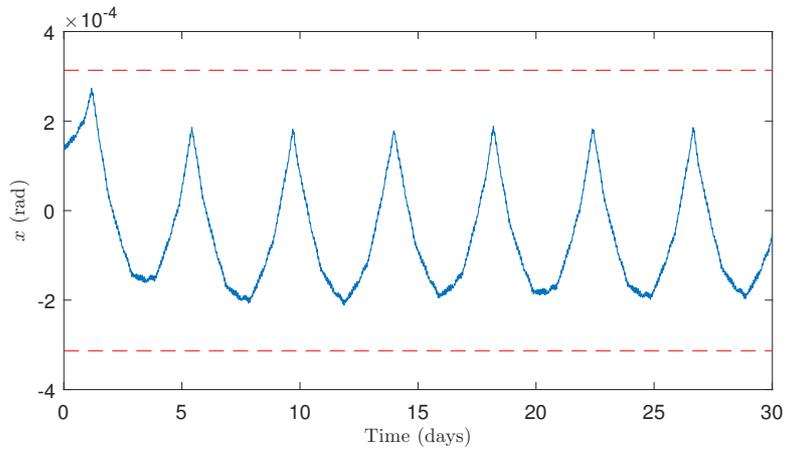}
\caption{Tracking error $x(t)$ (solid) and error tolerance $\pm x_\text{lim}$ (dashed) .}
\label{fig1}
\end{figure}
\begin{figure}[t]
\centering
\psfrag{v}{\small$v$}
\psfrag{T}{\footnotesize{Time (days)}}
\includegraphics[width=0.835\linewidth]{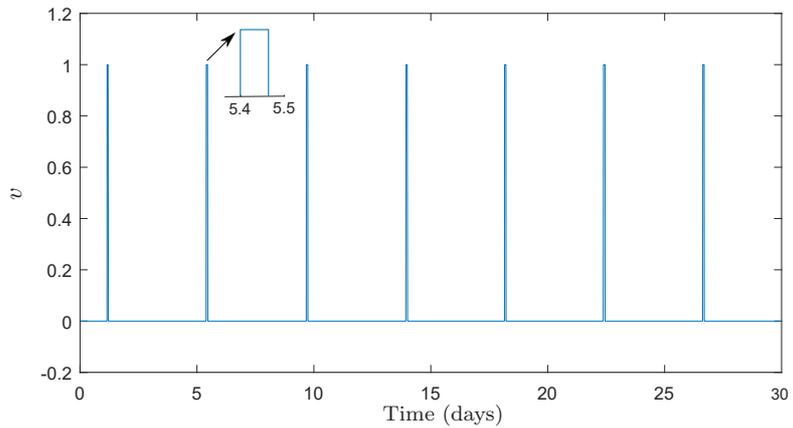}
\caption{Switching command $v(t)$.}
\label{fig2}
\end{figure}
The mission is simulated for 30 days. The tracking error $x(t)$ resulting from the simulation is reported in Fig.~\ref{fig1}. The corresponding switching command $v(t)$ is reported in Fig.~\ref{fig2}.
The controller is activated at $t\simeq$ 1 day (after the RLS filter transient has elapsed), in correspondence of the first peak in Fig \ref{fig1}. It can bee seen that a limit cycle with amplitude $2\,y_\text{lim}$ is established from the second input transition onwards, and that the error is kept well within the maximum allowed deviation $x_\text{lim}$ (represented by the dashed lines in Fig. \ref{fig1}). The HET engine is fired once about every 4 days, for a time interval of approximately 94 minutes. The duty cycle is equal to $D=0.0156$: such a small value is due to the fact that the atmospheric drag force is much smaller than the $10$ mN thrust force produced by the engine.

\begin{figure}[t]
\centering
\includegraphics[width=0.85\linewidth]{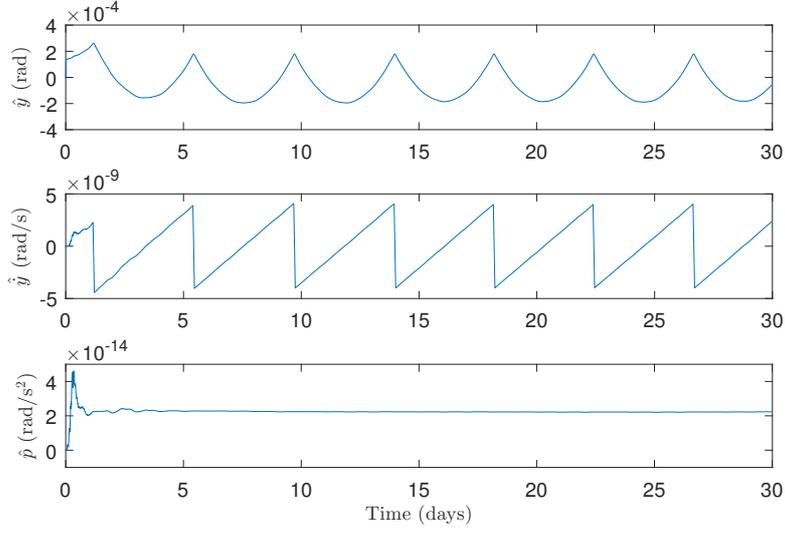}
\caption{Estimates $\hat{y}(t)$, $\hat{\dot{y}}(t)$ and $\hat{p}(t)$ returned by the RLS filter.}
\label{fig3}
\end{figure}
\begin{figure}[!t]
\centering
\includegraphics[width=0.845\linewidth]{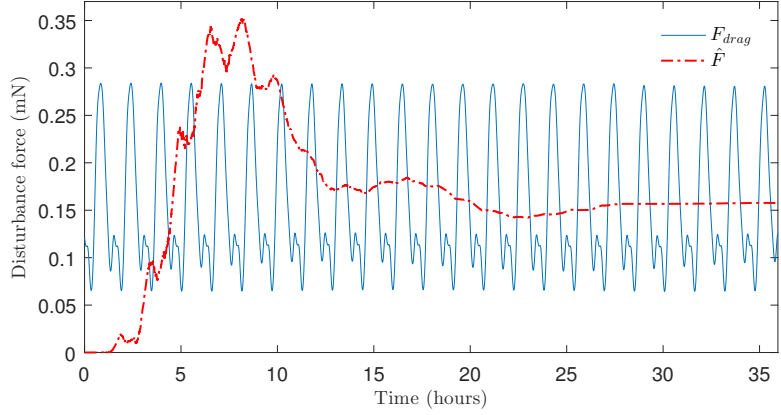}
\caption{Profiles of $F_\text{drag}$ (solid) and $\hat{F}$ (dash-dotted).}
\label{fig4}
\end{figure}

The signals $\hat{y}(t)$, $\hat{\dot{y}}(t)$ and $\hat{p}(t)$, estimated by the RLS filter, are shown in Fig.~\ref{fig3}. It can be noticed that the estimates reach a steady state in approximately one day. In particular, $\hat{p}(t)$ settles to a value which is very close to the average acceleration due to drag, as explained in Section \ref{sec3}. To better illustrate this fact, Figure \ref{fig4} compares the estimated disturbance force $\hat{F}=m_\text{sat}\, a^*\,\hat{p}(t)/(3\,r)$ with the absolute value $F_\text{drag}$ of the drag force returned by the  NRLMSISE-00 model, on the time interval $t\in[0,36]$ hours. The oscillations in $F_\text{drag}$ are due to changes in the atmospheric density induced by diurnal, latitudinal, and altitude variations. It is confirmed that the RLS filter is able to average out these effects, at steady-state. This is a key requirement for the implementation of the proposed control scheme.

\begin{figure}[!t]
\centering
\includegraphics[width=0.85\linewidth]{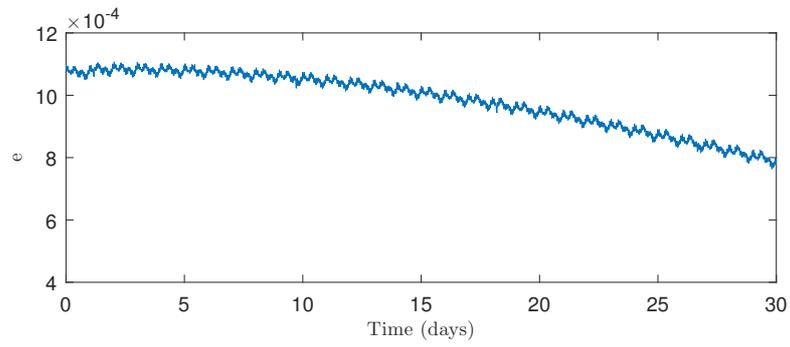}
\caption{Evolution of the orbital eccentricity $e$.}
\label{fig5}
\end{figure}

The evolution of the orbital eccentricity is reported in Fig. \ref{fig5}. As expected, the proposed thrusting strategy has a negligible impact on this element, which shows a slow decrease due to environmental perturbations. Hence, the orbit of the spacecraft remains approximately circular during the entire simulation interval.

\begin{figure}[!t]
\centering
\includegraphics[width=\linewidth]{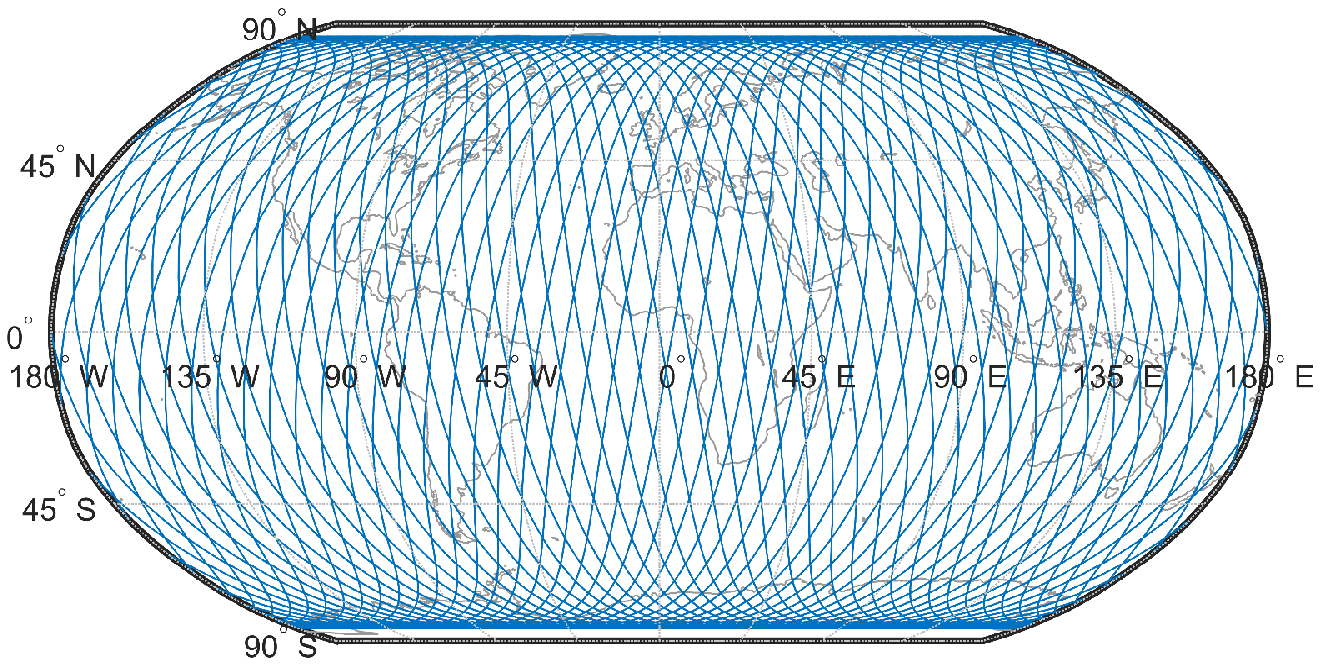}
\caption{Controlled groundtrack.}
\label{fig6}
\end{figure}

\begin{figure}[!t]
\centering
\includegraphics[width=\linewidth]{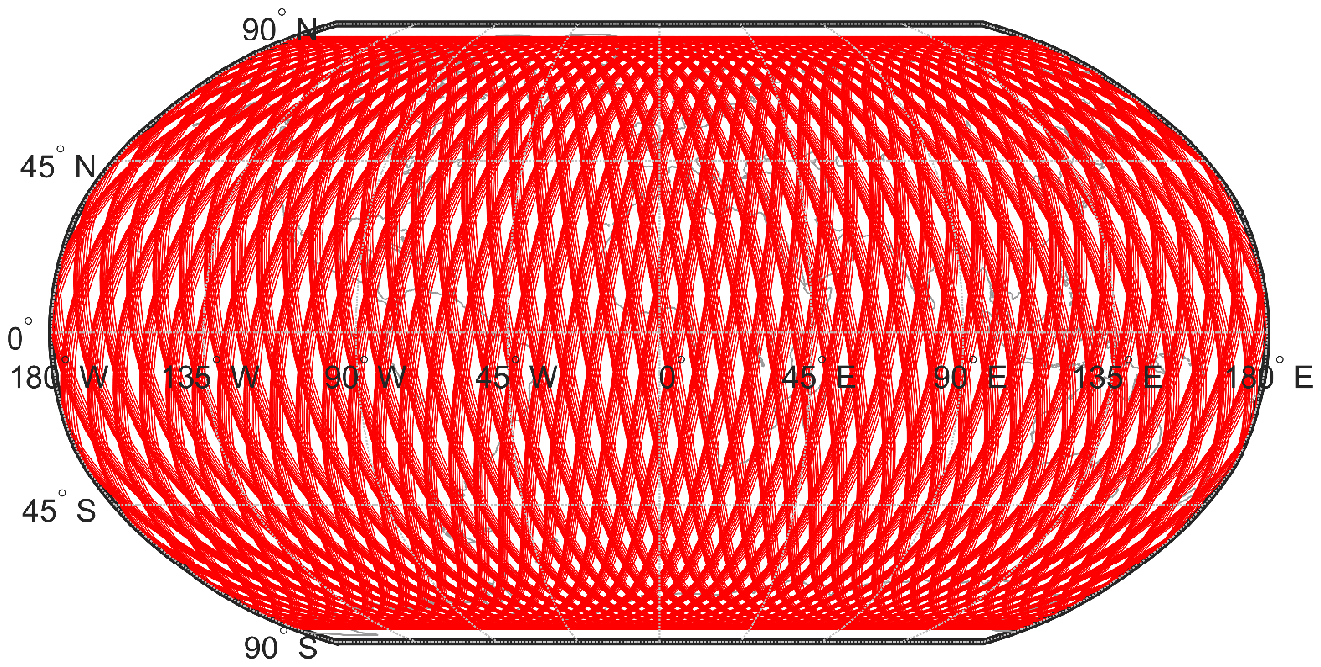}
\caption{Uncontrolled groundtrack.}
\label{fig7}
\end{figure}

Finally, the groundtrack profile resulting from the simulation has been compared to the uncontrolled one. As seen from Figs~\ref{fig6}-\ref{fig7}, the proposed method is able to compensate for the groundtrack drift due to perturbation effects, thus ensuring precise repeatability. It is worth stressing that the method requires neither a gravitational nor an atmospheric drag model: the controlled groundtrack automatically settles to the desired repeat condition, given only the parameter $r$ in~\eqref{trkerr} and the measurements provided by the GPS. Moreover, notice that one may consider updating the parameter $k$ in \eqref{kpar} on a periodic basis, should the engine acceleration deviate significantly from the nominal design value $u_\text{max}$.

\section{Conclusions}\label{sec6}
A simple and effective groundtrack maintenance strategy has been presented for low Earth orbiting satellites driven by low-thrust propulsion. The proposed adaptive control scheme consists of an hysteresis controller paired with a recursive least squares filter. It can be readily implemented within an autonomous guidance, navigation and control system.  The results of a simulation case study show that the desired repeat-groundtrack pattern is acquired successfully and then maintained consistently by the control system. Such type of technology may play a key role in a number of future scientific and commercial space missions equipped with electric propulsion.


%\phantomsection
%\addcontentsline{toc}{chapter}{References}
%\section*{References}
\bibliographystyle{aiaa-doi}
\bibliography{biblio}

\clearpage
\phantom{}
\authorbibliography[scale=0.45,imagewidth=2.2cm,wraplines=7,overhang=-5pt]{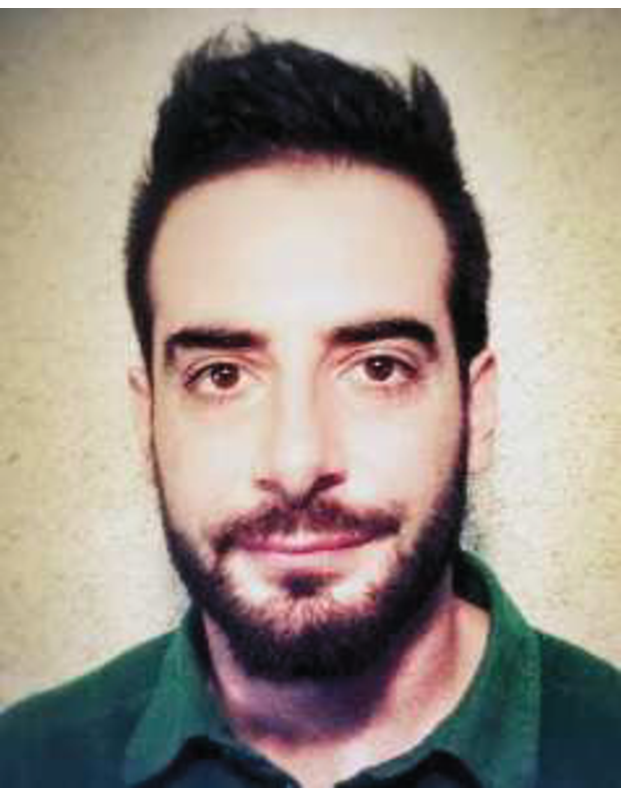}{}{\textbf{Mirko Leomanni}
was born in Siena, Italy, in 1983. He received the M.Sc. degree in Information Engineering in 2008, and the Ph.D. in Information Engineering and Science in 2015, both from the University of Siena. Since 2015, he is a research associate in automatic control and robotics at the University of Siena. His research interests include spacecraft dynamics and control, analysis of switching systems, optimization, and autonomous navigation.}

\authorbibliography[scale=0.64,imagewidth=2.2cm,wraplines=7,overhang=-5pt]{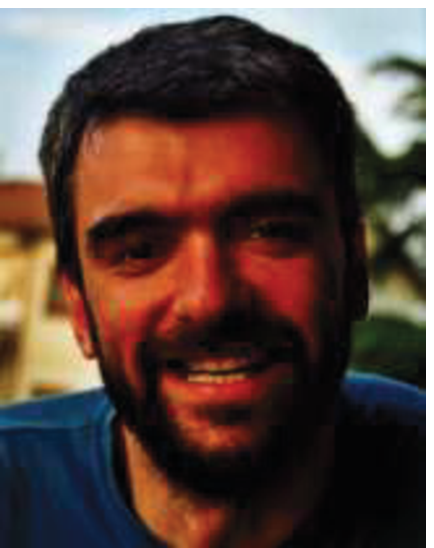}{}{\textbf{Andrea Garulli} was born in Bologna, Italy, in 1968. He received the Laurea in
Electronic Engineering from the  Universit\`a di Firenze in 1993, and the Ph.D. in System Engineering from the Universit\`a di Bologna in 1997. In 1996 he
joined the Universit\`a di Siena, where he is currently Professor of Control Systems. Since 2015, he is the director of the Dipartimento di Ingegneria dell'Informazione e Scienze Matematiche. He has been member of the Conference Editorial  Board of the IEEE Control Systems Society and Associate Editor of the IEEE Transactions on Automatic Control. He currently serves as Associate Editor for Automatica. He is author of more than 170 technical publications, co-author of the book \virg{Homogeneous Polynomial Forms for Robustness Analysis of Uncertain Systems} (Springer, 2009) and co-editor of  the books \virg{Robustness in Identification and Control} (Springer, 1999), and \virg{Positive Polynomials in Control} (Springer, 2005). His present research interests include system identification, robust estimation and filtering, robust control, mobile
robotics, autonomous navigation and aerospace systems}

\authorbibliography[scale=0.5,imagewidth=2.2cm,wraplines=7,overhang=-5pt]{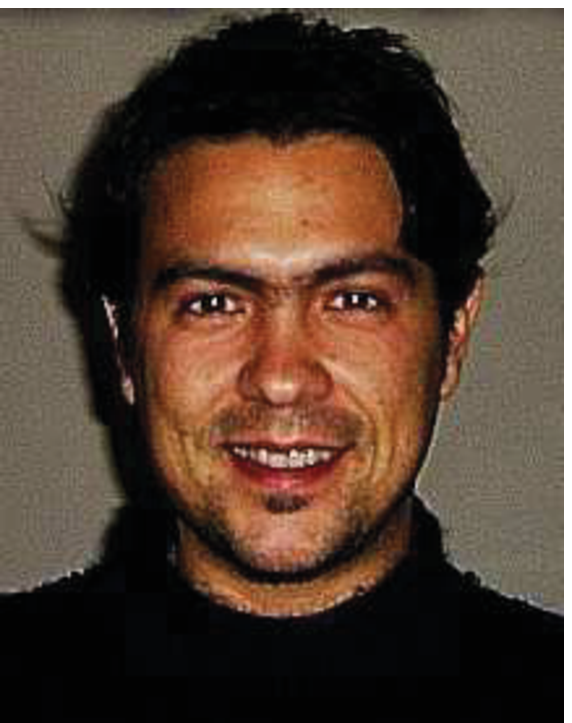}{}{\textbf{Antonio Giannitrapani} was born in Salerno, Italy, in 1975. He received the Laurea degree in Information Engineering in 2000, and the Ph.D. in Control Systems Engineering in 2004, both from the University of Siena. In 2005 he
joined the Dipartimento di Ingegneria dell’Informazione e Scienze Matematiche of the same university, where he is currently Assistant Professor. His research interests include localization and map building for mobile robots,
motion coordination of teams of autonomous agents and attitude control systems of satellites.}

\authorbibliography[scale=0.64,imagewidth=2.2cm,wraplines=7,overhang=-5pt]{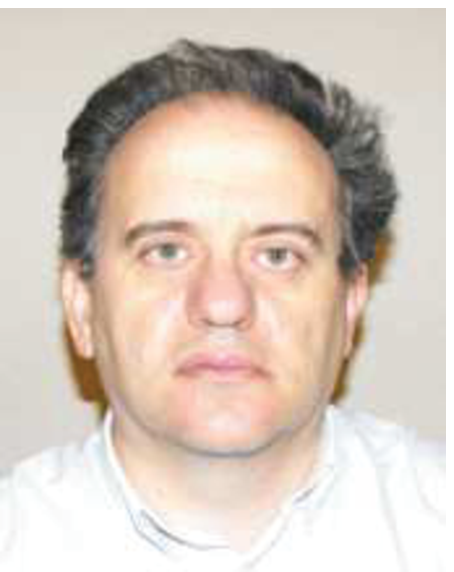}{}{\textbf{Fabrizio Scortecci} received a MS in Aerospace Engineering at the
Universit\`a di Pisa in 1990. Since his graduation he worked as a Researcher
and then as a Project Manager in various theoretical and experimental
projects related to electric satellite propulsion, aerothermodynamics
and spacecraft systems. During the year 2000 he joined AEROSPAZIO
Tecnologie s.r.l. working as Senior Scientist and Manager on programs
related to on-orbit application of electric propulsion.}

\end{document}